\documentclass[twocolumn]{article}

\usepackage{lipsum} 
\usepackage{graphicx} 
\usepackage{caption} 
\usepackage{amsmath} 
\usepackage{hyperref} 
\usepackage{authblk} 
\usepackage{titlesec} 
\usepackage{marginnote} 
\usepackage{orcidlink}
\usepackage{subcaption} 
\date{}

\usepackage[margin=1in]{geometry} 

\title{\huge{\textbf{Proactive Emotion Tracker: AI-Driven Continuous Mood and Emotion Monitoring}}}
\author[1]{Mohammad Asif\orcidlink{0000-0002-9517-6716}\thanks{email:, \texttt{pse2017001@iiita.ac.in}}}
\author[2]{Sudhakar Mishra\orcidlink{0000-0002-3748-7153}\thanks{email:, \texttt{sudhakarm@iitk.ac.in}}}
\author[1]{Ankush Sonker}
\author[1]{Sanidhya Gupta}
\author[1]{Somesh Kumar Maurya}
\author[1]{Uma Shanker Tiwary\orcidlink{0000-0001-7206-9013}}

\affil[1]{Indian Institute of Information Technology, Allahabad, India}
\affil[2]{Indian Institute of Technology, Kanpur, India}

\renewcommand{\thesection}{\Roman{section}} 

\titleformat{\section}[block]{\centering\large\bfseries}{\thesection.}{0.5em}{}

\begin{document}

\maketitle

\begin{abstract}
\textbf{This research project aims to tackle the growing mental health challenges in today's digital age. It employs a modified pre-trained BERT model to detect depressive text within social media and users' web browsing data, achieving an impressive 93\% test accuracy. Simultaneously, the project aims to incorporate physiological signals from wearable devices, such as smartwatches and EEG sensors, to provide long-term tracking and prognosis of mood disorders and emotional states. This comprehensive approach holds promise for enhancing early detection of depression and advancing overall mental health outcomes.}
\end{abstract}

\section{Introduction}
In today's digital age, maintaining good mental health has become a challenging task. The rates of depression, anxiety, and stress have been increasing at an alarming rate. However, diagnosing depression early on is not easy due to a lack of knowledge and ambiguous symptoms that are difficult to measure. Additionally, people who are suffering from these symptoms may hesitate to share them with others, such as experts or friends. This work is an effort to digitally track the mood swings and emotional patterns of any individual by analysing-
\begin{enumerate}
    \item Individual’s social media text data and browsing data of web search history; and
    \item Physiological signals captured through wearable devices (such as smartwatches or EEG sensors) on certain occasions.
\end{enumerate}

\section{Problem Statement}
In this work, our goal is to tackle depression and various mood disorders. Depression has been a major threat to society, so we have presented a solution for identifying such disorders. It is evident that Social Media is an emerging platform where people generally report or confess their mental state. Currently, it has become apparent that many people experiencing mood disorders are turning to browsing the internet and using search engines to find solutions to their symptoms. Individuals from various parts of the globe have adopted this practice. To tackle this, we will be using the browser history of the person as test data. We will create an interface which can be integrated into the browser, which can analyse browser history in the background using pre-trained ML models\cite{10.1007/s10772-023-10021-4, 10.1007/s11042-022-12172-z} and predict the various moods. It can benefit individuals in maintaining a real-time record of their moods and mental state.

The second part of the project will consist of building a framework that will work with emotion recognition\cite{asif2022emotion, asif2024deep, 10329776} using wearable devices (e.g., smartwatches, etc.), EEG and other physiological signals (heart rate, temperature) \cite{10101783}. The framework will take a prognosis of various mood disorders and/or emotional states by recognising and tracking emotions for the long term\cite{mishra2022cardiac,mishra2022dynamic}.

\section{Objectives}
\begin{enumerate}
    \item Improving the overall mental health and wellness of the people utilizing AI and machine learning.
    \item Investigating factors responsible for mood swings and emotion-related disorders.
    \item Real-time mood monitoring of moods/ mental states from social media data.
    \item Long-term emotion detection and tracking using EEG for various groups and cultures.
    \item Prognosis of various moods and emotional states.
\end{enumerate}

\section{Methodology}
\subsection{Approach}
\subsubsection{Dataset Creation and Model Training:} Extract the latest tweets from Twitter containing a wide and dynamic spectrum of Twitter Tweets datasets from Kaggle dealing in the context of depressive and non-depressive texts. This would be acting as the source of training data to train the model. Post extraction, pre-process the data using tokenization, stemming, stop-words and TF-IDF methodologies to extract only texts and remove URLs, hash-tags and useless words from the dataset. Train an ML model using this above-created dataset.

\subsubsection{Create a Script to Extract User Browser History:} We would be creating a script or a program which keeps running and executing in the background and fetches the browser history of the user.

\begin{figure}[ht]
  \centering
  \includegraphics[width=\columnwidth]{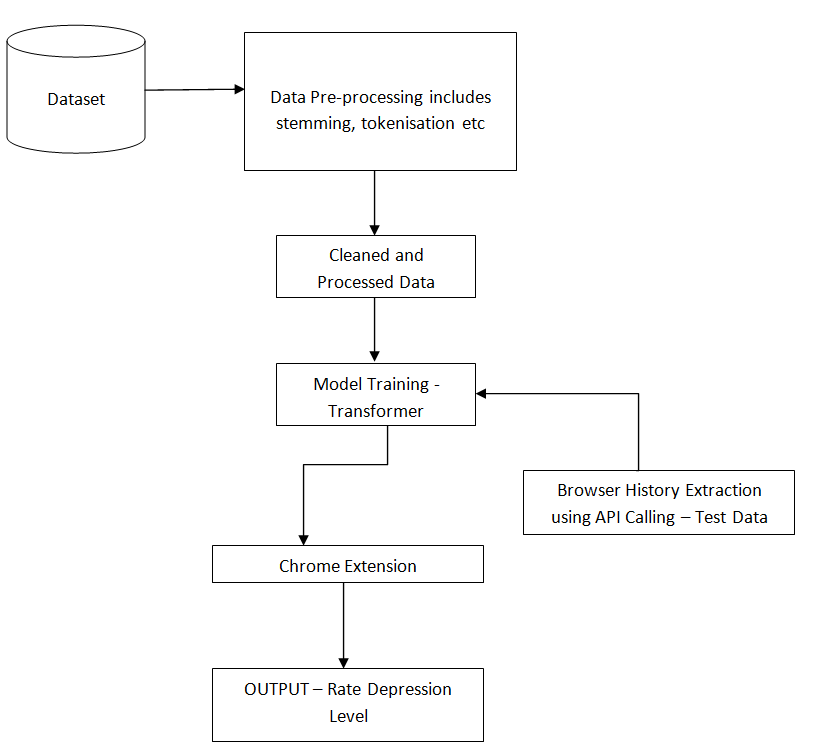}
  \caption{An overall view of the work for social media and browsing history analysis for mood disorders detection.}
  \label{fig:overall}
\end{figure}

The complete procedure flow is depicted in Fig.\ref{fig:overall}.

\subsection{Datasets}
We used the standard dataset from Kaggle\cite{suicide_watch_dataset}. It contains a wide range of flexible, dynamic, and a wide variety of texts dealing with depressive and non-depressive texts.  Also, we have used the Twitter Tweets APIs dataset and mixed both datasets to create a bigger and more flexible dataset. The tweets collected provide a broad and dynamic spectrum of depressive and non-depressive texts.

\subsection{Model Architecture}
We introduced a technique where we are using the BERT Model, which is basically one of the standard models in the Transformers series. Also, in order to test the model, we have extensively created a browser-based extension that extracts the data from the user's browser and analyses all the history data extracted together. This data is then tested for the detection of the level of the seriousness of depression.

\begin{figure}[ht]
  \centering
  \includegraphics[width=\columnwidth]{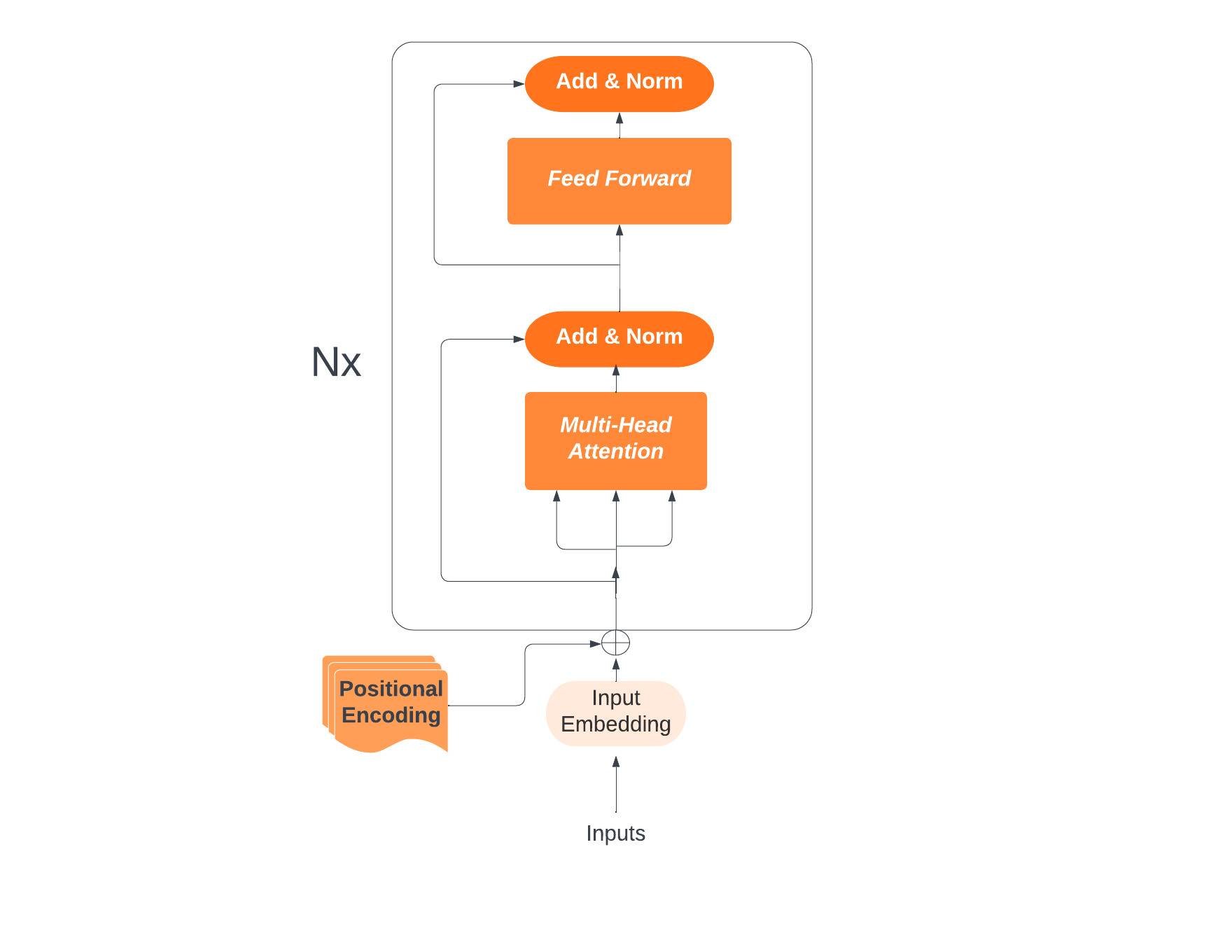}
  \caption{Proposed model architecture of BERT Transformer.}
  \label{fig:model}
\end{figure}

We used transfer learning on the BERT transformer model to predict whether a given sentence is depressive or not. The Bidirectional Encoder Representations from Transformers (BERT) model (see Fig.\ref{fig:model}) is a breakthrough in natural language processing (NLP) that has revolutionized various NLP tasks, including language understanding, sentiment analysis, question answering, and text classification. The fundamental concept behind BERT is pre-training and fine-tuning. BERT is pre-trained on two tasks: Masked Language Modelling (MLM), in which it learns to predict missing words in sentences and Next Sentence Prediction (NSP), in which it learns to predict whether one sentence is the subsequent sentence to another sentence in the original document.

\subsection{Fine Tuning}
With our understanding of the Transformers, we realised that the initial layers of the model adapt to recognise features that are low-level to the context and can be easily transferred if the basic context of both problems is the same. However, the later layers of the model adapt more towards high-level features and are very specific to the details of the problem. So, in our approach, we unfreezed the last layer, i.e., the pooler layer and last encoder block (multi-head attention and feedforward net) of the BERT and trained it over the dataset as well.

\section{Results}
\begin{figure}[ht!]
  \centering
  \begin{subfigure}{0.49\textwidth}
    \includegraphics[width=\textwidth]{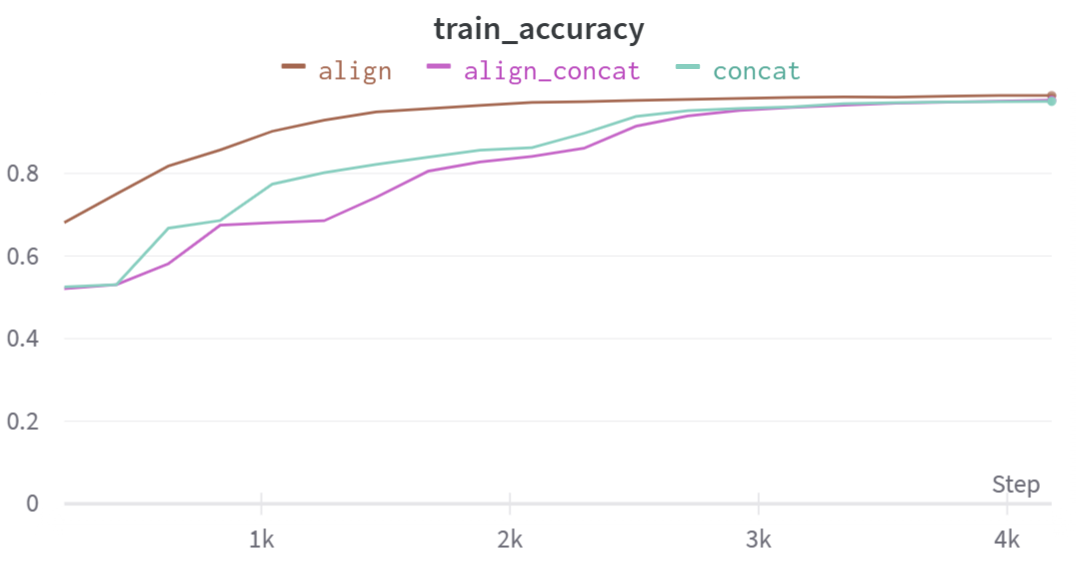}
    \label{fig:subfig-a}
  \end{subfigure}
  \hfill
  \begin{subfigure}{0.49\textwidth}
    \includegraphics[width=\textwidth]{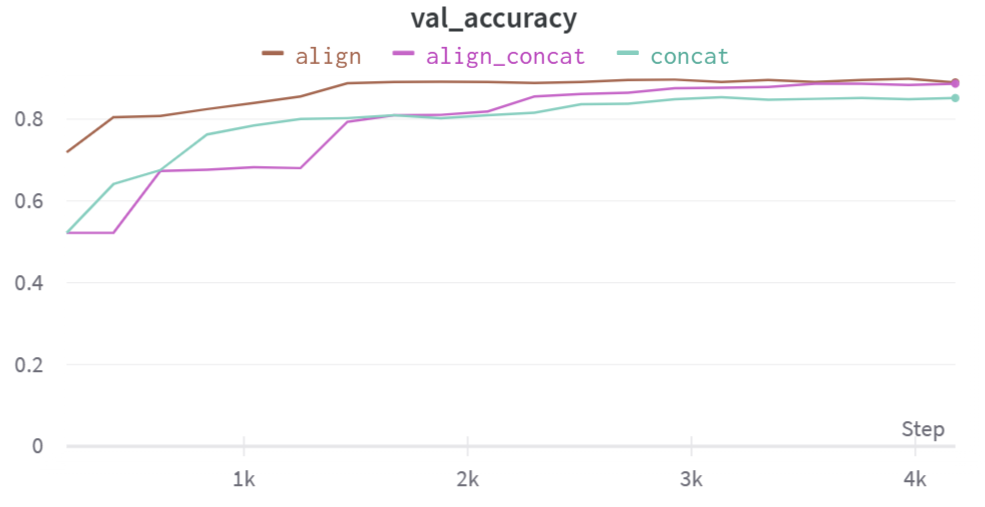}
    \label{fig:subfig-b}
  \end{subfigure}
  
  \vspace{1em}
  
  \begin{subfigure}{0.49\textwidth}
    \includegraphics[width=\textwidth]{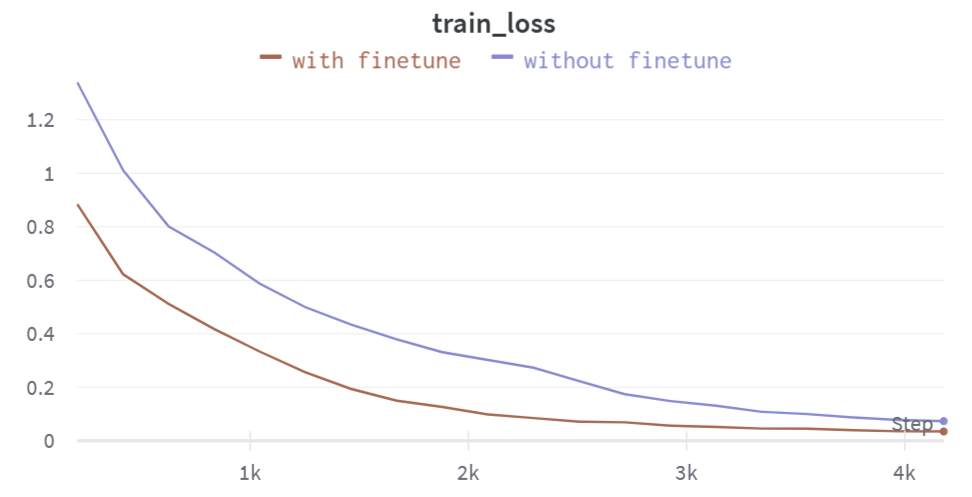}
    \label{fig:subfig-c}
  \end{subfigure}
  \hfill
  \begin{subfigure}{0.49\textwidth}
    \includegraphics[width=\textwidth]{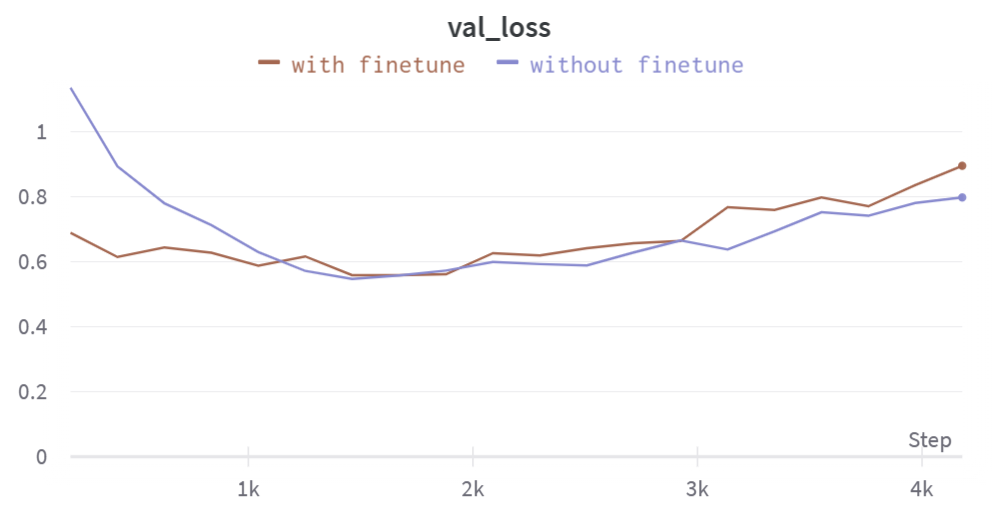}
    \label{fig:subfig-d}
  \end{subfigure}
  
  \caption{Accuracy and Loss Graphs}
  \label{fig:acc_loss}
\end{figure}

By training this modified pre-trained BERT (re-initialising pooler layer and the last encoder layer) on our dataset, we got far better performance: 93\% test accuracy, whereas, with direct fine-tuning, we got 82.47\% accuracy. Please see Fig. \ref{fig:acc_loss} for more insights into the results.

\section{Conclusion and Future Work}
In conclusion, using our method, we are getting promising results. We think our model can work on real-time data from the social-media platform. One of the possible difficulties which is very immediate in our mind is the privacy concern of our participants. We will try to convince them that their data is not being transferred to any server, and upon their consent only, we will use their data to make any kind of prediction.

In our earlier research\cite{mishra2022dynamic}, we found that there are certain hub EEG electrodes that play a major role in emotion processing. We plan to customize the already available EEG bands with the targeted electrode locations\cite{mishra2022dynamic}. Although this option is less accessible to everyone (due to cost factors, maybe), the analysis of brain waves on a set of targeted electrodes will robustify our findings from the social media data. 
As an alternative low-cost wearable, we can use ECG sensors or data from smartwatches, etc., to monitor and predict emotional states as per our findings in\cite{mishra2022cardiac}.

\bibliographystyle{unsrt}
\bibliography{biblio}  

\end{document}